\documentclass[conference,10pt]{IEEEtran}
\usepackage{setspace}
\usepackage{array}
\usepackage{xcolor}
\usepackage{graphicx}
\usepackage[cmex10]{amsmath}
\usepackage{amsthm}
\usepackage{mathrsfs}
\usepackage{txfonts}
\usepackage{cite}
\usepackage{color}
\usepackage{subfig}
\usepackage{longtable}
\usepackage{multirow}
\usepackage{bm}
\DeclareMathOperator*{\Res}{Res}

\ifCLASSINFOpdf
  
\else
  
\fi

\begin{document}
\newtheorem{theorem}{Theorem}[section]
\newtheorem{proposition}{Proposition}[section]
\newtheorem{lemma}{Lemma}[section]
\newtheorem{corollary}[theorem]{Corollary}
\newtheorem{example}{Example}[section]
\newtheorem{definition}{Definition}[section]
\newcommand{\BEQA}{\begin{eqnarray}}
\newcommand{\EEQA}{\end{eqnarray}}
\newcommand{\define}{\stackrel{\triangle}{=}}
\newcommand{\eqtext}[1]{\ensuremath{\stackrel{\text{#1}}{=}}}
\newcommand*{\QEDA}{\hfill\ensuremath{\blacksquare}}%
\newcommand*{\QEDB}{\hfill\ensuremath{\square}}%
\bibliographystyle{IEEEtran}

\theoremstyle{remark}
\newtheorem{rem}{Remark}
\providecommand{\abs}[1]{\lvert#1\rvert}
\providecommand{\res}[1]{\Res\displaylimits_{#1}} 
\providecommand{\norm}[1]{\lVert#1\rVert}
\providecommand{\mtx}[1]{\mathbf{#1}}

\providecommand{\pr}[1]{\ensuremath{\Pr\left(#1\right)}}
\providecommand{\qfunc}[1]{\ensuremath{Q\left(#1\right)}}
\providecommand{\sbrak}[1]{\ensuremath{{}\left[#1\right]}}
\providecommand{\lsbrak}[1]{\ensuremath{{}\left[#1\right.}}
\providecommand{\rsbrak}[1]{\ensuremath{{}\left.#1\right]}}
\providecommand{\brak}[1]{\ensuremath{\left(#1\right)}}
\providecommand{\lbrak}[1]{\ensuremath{\left(#1\right.}}
\providecommand{\rbrak}[1]{\ensuremath{\left.#1\right)}}
\providecommand{\cbrak}[1]{\ensuremath{\left\{#1\right\}}}
\providecommand{\lcbrak}[1]{\ensuremath{\left\{#1\right.}}
\providecommand{\rcbrak}[1]{\ensuremath{\left.#1\right\}}}

\title{Closed-form Approximations for Coverage and Rate in a Multi-tier Heterogeneous Network in Nakagami-m Fading}
%
%
%

\author{G. V. S. S. Praneeth Varma, G. V. V. Sharma, {\em Member, IEEE}, and A. Kumar, {\em Member, IEEE}
\thanks{
G. V. S. S. Praneeth Varma, G. V. V. Sharma,
and A. Kumar are with the Department of Electrical
Engineering, Indian Institute of Technology Hyderabad,
Kandi, Sangareddy, Telangana 502285, India
(e-mail: {\tt \{ee14resch11007, gadepall, abhinavkumar\}@iith.ac.in} ).}}

\maketitle

\begin{abstract}
In this paper, we consider the downlink in a K-tier heterogeneous network in the presence of Nakagami-m fading and noise. For such a system,
we derive closed-form approximations
of coverage probability and average rate achievable.
A piece-wise linear approximation is employed in obtaining the simplified expressions. The proposed results are verified numerically through simulations. A comparison with existing work shows that the proposed work is a good approximation. 
\end{abstract}

\IEEEpeerreviewmaketitle


\section{Introduction}
The demand for higher data rates in cellular networks
has lead to the deployment of small cells along with
the macro base stations (BS) resulting
in multi-tier heterogeneous networks (HetNets).
The irregular BS deployment in such HetNets
can be suitably modeled as a
Poisson point process (PPP) as shown in 
\cite{ktier_downlink}.
In presence of Rayleigh fading,
a stochastic geometry based
approach has been used to determine
coverage probability and average rate achievable
for multi-tier HetNets in \cite{ktier_downlink}.
In the absence of simple analytical
expressions of coverage and rate,
ordering results for various transmission techniques
in multi-antenna HetNets with Rayleigh fading
for an interference limited scenario
have been presented in \cite{downlink_mimo}.
 In \cite{Renzo_meijer}, average rate for generalized fading channels were derived using an MGF approach. An  expression for the coverage
probability was derived in semi-closed form for the dual branch in multi-antenna single tier network in \cite{dual_branch}. Coverage and rate were derived using the gil-peleaz inversion formula in \cite{Renzo_gil_pelaez}.  Analytical results in the above literature were expressed in either  single or two fold integrals. A closed form expression for outage in the presence of rayleigh fading for a single tier network in terms of a toeplitz matrix was obtained in \cite{khaled_toeplitz}. This problem was extended and solved for MIMO in \cite{khaled_toeplitz_mimo}, where the decision variable involved Gamma random variables.  However, both the above papers focused on interference limited systems. 

Thus, simplified analytical expressions
for coverage and rate in the presence of
Nakagami-m fading for both noise and interference
limited scenarios are required. This is the motivation of this work.  The coverage expressions obtained in our work are approximate, but the approach is extremely simple and the results quite accurate. 
The system model considered in this paper is presented next.
%
\section{System model}
 We consider a $K$-tier HetNet
such that each tier $i$'s BSs are
distributed according to a PPP $\Phi_i$
of density $\lambda_i$.
The BSs in a tier $i$
have same transmit power $P_i$
and signal-to-interference-plus-noise ratio
(SINR) threshold $\beta_i$.
The path loss from a location $x_i$
to the origin is defined as $L(x_i)=||x_i||^{-\alpha}$.
The SINR for a typical user equipment (UE)
at the origin from a
BS located at $x_i$, in the tier $i$, is
 \begin{equation}
 SINR(x_i)=\frac{P_ih_{x_i}||x_i||^{-\alpha}}{\sum_{j=1}^K\sum_{x\in \Phi_j\backslash x_i}P_jh_{x}||x||^{-\alpha}+\sigma^2} \, ,
 \label{SINR}
 \end{equation}
where,
$h_{x_i}$ is the fading power between the UE
and the BS at location ${x_i}$,
and $\sigma^2$ is the noise power.
The fading power from all the BSs
is assumed to be independently distributed
such that  $h_{x_i} \sim \Gamma (M_i,1)$
has a Gamma distribution, i.e.,
Nakagami fading.
Further, for every tier $i$,
$h_{x_i}$'s are independent and identically distributed (i.i.d). 
A coverage event for the typical user $
\mathbf{C}(\{\beta_i\} )$
is defined for the set $\{\beta_i\} $ as
\begin{equation}
\mathbf{C}(\{\beta_i\} )
\triangleq \cup_{i \in K, x_i \in \Phi_i} (SINR(x_i) > \beta_i ) \, .
\end{equation}
Then, from \cite{ktier_downlink},
the coverage probability $P_c$ for
Nakagami-$m$ fading is expressed as
\begin{equation}
P_c= \mathbb{P}\left( \mathbf{C}(\{\beta_i\})\right)=\mathbb{P}\left(\ \cup_{i \in K, x_i \in \Phi_i} (SINR(x_i) > \beta_i) \right)  \, .
\label{CP}
\end{equation}
Under the assumption $\beta_i> 1$, i.e.,
at most one BS in the entire
network can provide SINR greater than the required threshold and
using (\ref{SINR}),
(\ref{CP}) simplifies to \cite{ktier_downlink} 
\begin{equation}
\label{general_cov}
P_c=\sum_{i=1}^K \lambda_i \int_{R^2}\mathbb{P}\left(\frac{P_ih_{x_i}L(x_i)}{I_{x_i}+\sigma^2} > \beta_i\right)dx_i \, .
\end{equation}
\setcounter{equation}{8}
\begin{figure*}[!t]
\small
{
\begin{eqnarray}
\mathcal{I}_i &=& \sum_{k=0}^{M_i-1} \frac{1}{k!}\sum_{l=0}^k\binom{k}{l}\frac{(\sigma^2)^{k-l}(-1)^l}{A^{\frac{\alpha}{2}(k- l)+1}}\sum_{r=0}^l (-1)^rB_{l,r}(D_1,D_2,\dots,D_{l-r+1}) 
\lsbrak{c\left\{\gamma\left( r+\frac{\alpha}{2}(k- l)+1,\frac{A}{(\sigma^2)^{2/\alpha}}x_2 \right) \right.
\left.-\gamma\left(r+\frac{\alpha}{2}(k- l)+1,\frac{A}{(\sigma^2)^{2/\alpha}}x_1 \right) \right\}} 
\nonumber  \\
& &+\gamma\left(r+\frac{\alpha}{2}(k- l)+1,\frac{A}{(\sigma^2)^{2/\alpha}}x_1\right)  
+ \frac{(\sigma^2)^{2/\alpha}}{A}m\left\{\gamma\left( r+\frac{\alpha}{2}(k- l)+2,\frac{A}{(\sigma^2)^{2/\alpha}}x_2 \right) \right.
\rsbrak{\left.-\gamma\left(r+\frac{\alpha}{2}(k- l)+2,\frac{A}{(\sigma^2)^{2/\alpha}}x_1 \right) \right\} }  \label{eqn_Ii} 
\end{eqnarray}
 }
\hrulefill
\end{figure*}
\setcounter{equation}{4}
Given that the user is in coverage,
the average rate achievable
for Nakagami-m fading is expressed in \cite{ktier_downlink} as
\begin{equation}
R = \mathbb{E} \left[ log\left( 1 +\left.\underset{x\in \cup \Phi_i}{max} \, (SINR(x))\right|\mathbf{C}(\{\beta_i\}) \right)\right] \, , \nonumber
\end{equation}
which simplifies to
\begin{equation}
\label{rate_integral}
R = \int_{0}^{\infty}\frac{\mathbb{P}(X>y|\mathbf{C}(\{\beta_i\}))}{1+y}dy \, ,
\end{equation}
where, 
\begin{equation}
\label{ratio_Pc}
\mathbb{P}(X>y|\mathbf{C}(\{\beta_i\})) = \frac{\mathbb{P}(\mathbf{C}(\{max(y,\beta_i)\}))}{\mathbb{P}(\mathbf{C}(\{\beta_i\}))}.
\end{equation}
Next, we present the main results of this paper.

 \section{Coverage Probability}
%

A commonly encountered integral 
in the coverage analysis
using stochastic geometry based approach, as in \cite{ktier_downlink} \cite{downlink_mimo}, is
\begin{equation}
\int_{0}^{\infty} e^{-Vt-Ut^{\frac{\alpha}{2}}} t^\frac{n}{2}\,dt \, .
\label{eq5}
\end{equation}
For approximating (\ref{eq5}),
we present the following theorem.
\begin{theorem}
\label{pl_integral_thm}
For $\alpha > 2$,  $U>0$, and $V>0$,
\begin{multline}
\int_{0}^{\infty} e^{-Vt-Ut^{\frac{\alpha}{2}}} t^\frac{n}{2}\,dt
\approx \frac{1}{V^{\frac{n+2}{2}}}\lsbrak{\gamma\left(n/2+1,\frac{V}{U^{2/\alpha}}x_1\right) }
\\
+c\left\{\gamma\left( n/2+1,\frac{V}{U^{2/\alpha}}x_2 \right) -\gamma\left(n/2+1,\frac{V}{U^{2/\alpha}}x_1 \right) \right\} \\
 +\rsbrak{\frac{U^{2/\alpha}}{V}m\left\{\gamma\left( n/2+2,\frac{V}{U^{2/\alpha}}x_2 \right) -\gamma\left(n/2+2,\frac{V}{U^{2/\alpha}}x_1 \right) \right\} } \, ,%
\label{pl_integral_eqn}
\end{multline}
where, $m = -\frac{\alpha}{2} \brak{1-\frac{2}{\alpha}}^{1-\frac{2}{\alpha}}e^{-\brak{1-\frac{2}{\alpha}}}$,
$c= \frac{\alpha}{2}e^{-(1-\frac{2}{\alpha})}$,
$x_1=\frac{1-c}{m}$, $x_2=\frac{-c}{m}$, and 
$\gamma\brak{\cdot,\cdot}$ is the lower incomplete gamma function \cite[8.350]{table_integrals}.

\proof See Appendix \ref{lemma5_app}. \hfill \rule{1.5ex}{1.5ex}
\end{theorem}

Next, we present our main result for the
coverage probability in a K-tier HetNet. 

\addtocounter{equation}{1}
\begin{theorem}
\label{nak_cov_thm}
The coverage probability of K-tier HetNet in Nakagami fading, when $\beta_i > 1$, is
\begin{equation}
\label{nak_cov_eqn}
 P_c=\sum_{i=1}^K \pi\lambda_iP_i^{2/\alpha}\beta_i^{-2/\alpha} \mathcal{I}_i \, ,
 \end{equation}
where, $\mathcal{I}_i$ is as expressed in \eqref{eqn_Ii} with the variables $D_t = \prod_{q=0}^{t-1}\brak{\frac{2}{\alpha}-q}$ , \, $\binom{k}{l} = \frac{k!}{l! (k-l)!}$,
$A = \sum_{m=1}^K\lambda_m(P_m)^{2/\alpha} 
 \sum_{p=1}^{M_m}\binom{M_m}{p} \frac{2\pi}{\alpha}\mathcal{B}(M_m-p+2/\alpha,p-2/\alpha)$,
  \\ where, $\mathcal{B}\brak{\cdot,\cdot}$ is the Beta function \cite[8.380]{table_integrals} and the Bell polynomial is defined as
 \begin{equation}
 B_{l,r}(x_1,x_2,\dots,x_{l-r+1})
=\sum\frac{l!}{j_1!j_2!\dots j_{l-r+1}!}\prod_{t=1}^{l-r+1}\left(\frac{x_t}{t!}\right)^{j_t} 
\end{equation}
summation is over all $j$'s such that
$j_1+j_2+\dots+j_{l-r+1}=r $
and
 $j_1+2j_2+\dots+(l-r+1)j_{l-r+1}=l$ \, .
\proof See Appendix \ref{coverage_app}.
 \hfill \rule{1.5ex}{1.5ex}
\end{theorem}

\begin{corollary}
In the presence of Rayleigh fading, i.e., $M_i=1$, the coverage probability of a typical UE is
\begin{align}
\label{coverage_corollary}
P_c &=\sum_{i=1}^{K}\frac{\pi\lambda_iP_i^{2/\alpha}\beta_i^{-2/\alpha}}{V} \left[\left(1-e^{-\frac{V}{U^{2/\alpha}}x_1}\right) +c\left(e^{-\frac{V}{U^{2/\alpha}}x_1}-e^{-\frac{V}{U^{2/\alpha}}x_2}\right)\right. \nonumber\\
   &\quad +m \lcbrak{e^{-\frac{V}{U^{2/\alpha}}x_1}\brak{x_1+\frac{U^{2/\alpha}}{V}}}
    \left.\rcbrak{-e^{-\frac{V}{U^{2/\alpha}}x_2}\brak{x_2+\frac{U^{2/\alpha}}{V}}}  \right]
  \, ,
\end{align}
where, $V =  \frac{2\pi}{\alpha}\Gamma(2/\alpha)\Gamma(1-2/\alpha)\sum_{m=1}^K\lambda_mP_m^{2/\alpha} $\text{ and}
$U = \sigma^2$.
\end{corollary}
\proof
Substituting $M_i=1$ in (\ref{nak_cov_eqn}) and 
using $\gamma(1+z,x)=z!\left[1-e^{-x}\sum_{k=0}^z\frac{x^k}{k!}\right], \forall \, z \in \mathbb{Z}$
results in (\ref{coverage_corollary}).
\hfill \rule{1.5ex}{1.5ex}\\
Note that the result of Corollary III.3 is the closed form for the integral in  \cite[(2)]{ktier_downlink}.
Next, we present results for
the average rate achievable of a typical UE.
\section{Average Rate}
\begin{theorem}
\label{nak_rate_thm}
The average rate achievable of a typical UE in coverage
of a K-tier HetNet
in Nakagami fading, when $\beta_i > 1$, is
\begin{equation}
\label{nak_rate_eqn}
R=\frac{\sum_{i=1}^K \lambda_iP_i^{2/\alpha} \beta_i^{-2/\alpha} \mathcal{A}_i \mathcal{I}_i  }{\sum_{i=1}^K \lambda_iP_i^{2/\alpha}\beta_i^{-2/\alpha} \mathcal{I}_i } \, ,
 \end{equation}
 where, $\mathcal{I}_i$ is as in (\ref{eqn_Ii}),
\begin{align}
\mathcal{A}_i &= \ln\left(1+\beta_i\right)
+\frac{\alpha}{2} \ _2F_1\left(1,\frac{2}{\alpha},1+\frac{2}{\alpha};-\frac{1}{\beta_i}\right) \, ,
\end{align}
and $_2F_1(.)$ is the Gauss Hypergeometric function as
in \cite{encyclopedia_math}.
\end{theorem}
\proof See Appendix \ref{rate_app}. \hfill \rule{1.5ex}{1.5ex}
\begin{corollary}
In the presence of Rayleigh fading,
i.e., $M_i=1$,
the average rate achievable by a typical UE
in coverage is
\begin{equation}
\label{ray_rate_eqn}
R=\frac{\sum_{i=1}^K \lambda_iP_i^{2/\alpha} \beta_i^{-2/\alpha}\mathcal{A}_i }{\sum_{i=1}^K \lambda_iP_i^{2/\alpha}\beta_i^{-2/\alpha} } \, .
 \end{equation}
 \proof Substituting $M_i=1$ in (\ref{eqn_Ii}) results
in $I_i$ being a constant
with respect to (w.r.t.) $i$ which together with
(\ref{nak_rate_eqn}) results in (\ref{ray_rate_eqn}).
\hfill \rule{1.5ex}{1.5ex}
\end{corollary}

\begin{figure}[t]
\centering
\includegraphics[width=\columnwidth]{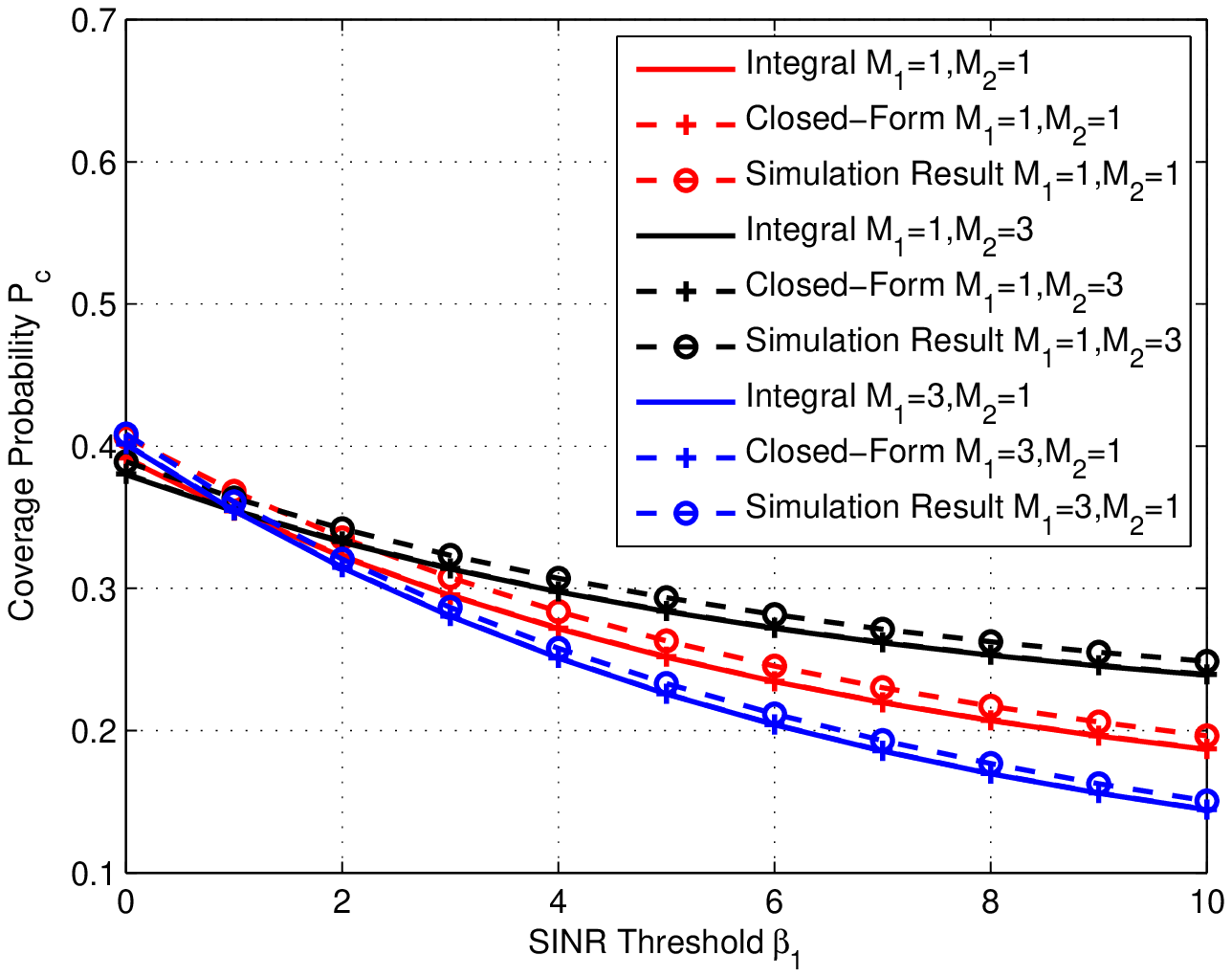}
\caption{Variation of $P_c$ w.r.t. $\beta_1$  in a 2-tier HetNet $(K=2,\alpha=3,P_1=25P_2,\lambda_2=5\lambda_1,\beta_2=1	dB)$}
\label{fig_cov_ray}
\end{figure}

 \section{Numerical Results}
  We consider the simulation setup as in \cite{ktier_downlink}, a two-tier HetNet consisting of macro BSs
  and small cells.
  We performed Monte Carlo simulations in MATLAB
  to obtain the simulation results which are averaged over $10^4$ location realizations each with
  $10^3$ channel realizations.
  We also numerically computed the integrals presented in \cite{ktier_downlink} and the proposed approximate
  expressions in MATLAB.  
  In Fig.~\ref{fig_cov_ray},
  we present the variation of $P_c$ w.r.t.
  $\beta_1$.
  
\begin{figure}[t]
\centering
\includegraphics[width=\columnwidth]{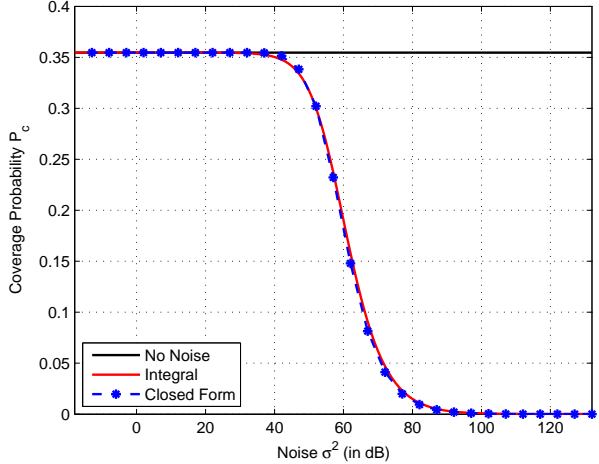}
\caption{Variation of $P_c$ w.r.t. $\sigma^2$ in a 2-tier HetNet $(K=2,\alpha=3,P_1=25P_2,\lambda_2=5\lambda_1,\beta_1=1dB,\beta_2=1dB)$}
\label{fig_cov_noise}
\end{figure}

 The curve generated using  (\ref{nak_cov_eqn}) matches closely with the existing
  result in \cite[(2)]{ktier_downlink} and the simulation results for various values of   the Nakagami parameters, $M_1$ and $M_2$.
  The variation of $P_c$ w.r.t.
  $\sigma^2$ in Rayleigh fading is presented in
  Fig.~\ref{fig_cov_noise} .  A good match is observed  even for 
 high values
  of $\sigma^2$, i.e., the noise limited regime.

\begin{figure}[t]
\centering
\includegraphics[width=\columnwidth]{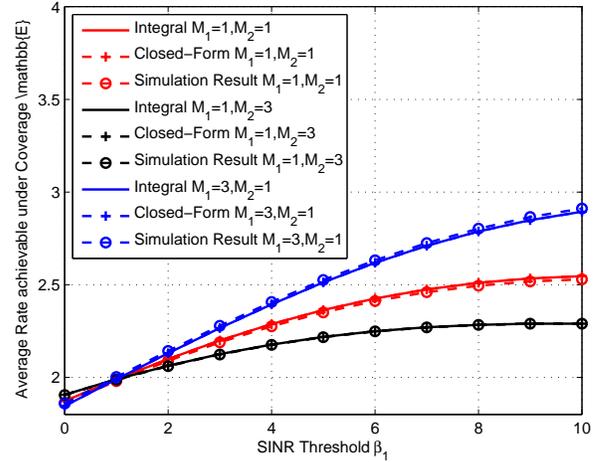}
\caption{Variation of $R$ w.r.t. $\beta_1$ when UE is in coverage $(K=2,\alpha=3,P_1=25P_2,\lambda_2=5\lambda_1,\beta_2=1dB)$}
\label{fig_rate_ray}
\end{figure}

  In Fig.~\ref{fig_rate_ray},
  the variation of  $R$ w.r.t. $\beta_1$ is presented. The curve generated using 
  (\ref{nak_rate_eqn}) matches closely
  with the integral in \cite[(7)]{ktier_downlink} and
  the simulation results as observed in Fig.~\ref{fig_rate_ray}.
  {\emph{Through extensive numerical computation
it was  verified that
the percentage loss in approximation of
$P_c$ expressed in (\ref{nak_cov_eqn}),
when using (\ref{pl_integral_eqn}),
is within $0.5\%$ for various values of $\alpha$, $M_i$, and
SINR}}.
  
%

%

 \section{Conclusion}
 We have proposed closed-form approximations
 for coverage probability and average rate achievable
 in a K-tier HetNet in the presence of noise and  Nakagami fading. Further, through simulation results
 we have shown that the proposed simplified expressions match closely with existing results.

\bibliography{IEEEabrv,hetnets}
\appendices
\section{Proof of Theorem III.1}
\label{lemma5_app}
\begin{figure}[t]
\centering
\includegraphics[width=\columnwidth]{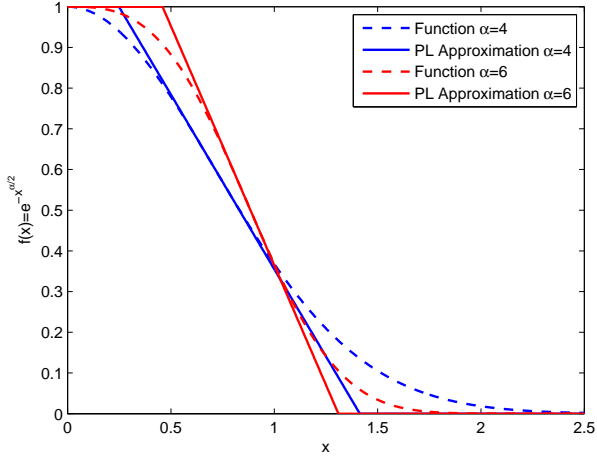}
\caption{Piecewise linear approximation}
\label{plapprox}
\end{figure}

The expression in \eqref{pl_integral_eqn}
can be simplified to
\begin{equation}
\int_{0}^{\infty} e^{-Ut^{\alpha/2}}e^{-Vt} t^\frac{n}{2}dt
=\frac{1}{U^{\frac{n+2}{\alpha}}}\int_{0}^{\infty} e^{-y^{\frac{\alpha}{2}}}e^{-\frac{V}{U^{2/\alpha}}y}y^{n/2} dy \, . \label{162}
\end{equation}
It is difficult to obtain an exact
closed form solution for
\eqref{162} for arbitrary $\alpha$.
From Fig. \ref{plapprox}, it can be seen that  $f(x) = e^{-x^{\alpha/2}}$
is monotonically decreasing w.r.t. $x$,
$f(x) \in (0,1] \, \forall x \geq 0$,
and has a single point of inflection.
Hence, for $f(x) = e^{-x^{\alpha/2}}$
a piece-wise linear
approximation (PLA) is given by
\begin{equation}
\label{pl_approx_fn}
e^{-x^{\alpha/2}} \approx
\left\{
	\begin{array}{ll}
		1  &  x \leq x_1 \, ,\\
		mx+c &  x_1	< x < x_2 \, ,\\	
		0 &  x_2 \leq x \, ,
	\end{array}
\right.
\end{equation}
where, $m$, $x_1$, $x_2$, and $c$
are unknown constants.
To obtain $m$, we calculate
the point of inflection
of $f(x)$ denoted by $x_0$ as follows,
\begin{equation}
\label{inflection_point}
\left.\frac{d^2(e^{-x^{\alpha/2}})}{dx^2}\right|_{x=x_0} =0 
\Rightarrow  x_0 = \left(1-\frac{2}{\alpha}\right)^{\frac{2}{\alpha}} \, .
\end{equation}
Using (\ref{inflection_point}),
the slope at $x_0$, i.e., $m$ is given by
\begin{align}
m=\sbrak{\frac{d\brak{e^{-x^{\frac{\alpha}{2}}}}}{dx}}_{x=x_0} 
= -\frac{\alpha}{2} \brak{1-\frac{2}{\alpha}}^{1-\frac{2}{\alpha}}e^{-\brak{1-\frac{2}{\alpha}}} \, .
\end{align}
From \eqref{pl_approx_fn},
we have 
\begin{eqnarray}
1 &= mx_1+c  \, , \label{eq161}
\\
0&=mx_2+c \nonumber \, ,
\\
e^{-x_0^{\alpha/2}} &= m x_0 +c \, ,  \nonumber
\end{eqnarray}
for $x$ equal to $x_1$, $x_2$, and $x_0$, respectively.
The three linear equations in \eqref{eq161}
can be jointly solved to obtain $x_1$, $x_2$,
and $c$ as given in \eqref{pl_integral_eqn}.

Substituting the approximation \eqref{pl_approx_fn} in \eqref{162}, we get
\begin{equation}
\frac{1}{U^{\frac{n+2}{\alpha}}}\left[\int_{0}^{x_1} y^{n/2}e^{-\frac{V}{U^{2/\alpha}}y} dy +\int_{x_1}^{x_2} (my+c)y^{n/2}e^{-\frac{V}{U^{2/\alpha}}y} \right] dy \nonumber \, ,
\end{equation}
which with transformation of variable results
in
\begin{multline}
\frac{1}{V^{\frac{n+2}{2}}} \! \lsbrak{ \! \int_{0}^{\frac{V}{U^{2/\alpha}}x_1} \!\!\!\! x^{n/2}e^{-x} dx \!+\! c\!\! \int_{\frac{V}{U^{2/\alpha}}x_1}^{\frac{V}{U^{2/\alpha}}x_2} \!\!\!\! x^{n/2}e^{-x} dx }  \!\\
+\! \rsbrak{\frac{U^{2/\alpha}}{V}m \! \! \int_{\frac{V}{U^{2/\alpha}}x_1}^{\frac{V}{U^{2/\alpha}}x_2} \!\!\!\! x^{n/2+1}e^{-x}dx }  \nonumber ,
\end{multline}
and using definition of 
lower incomplete gamma function \cite[8.350]{table_integrals}
results in \eqref{pl_integral_eqn}.
This completes the proof of Theorem \ref{pl_integral_thm}.
\hfill \rule{1.5ex}{1.5ex}

\section{Proof of Theorem III.2}
\label{coverage_app}
Given $h_{x_i} \sim \Gamma (M_i,1)$,
the conditional probability in \eqref{general_cov}
is
\begin{align}
\mathbb{P}\left(\left.\frac{P_ih_{x_i}L(x_i)}{I_{x_i}+\sigma^2} > \beta_i\right|I_{x_i} = I\right)
&=\int_{\frac{\beta_i(I+\sigma^2)}{P_iL(x_i)}}^{\infty}\frac{y^{M_i-1}e^{-y}}{\Gamma(M_i)}dy \, ,
\label{eq15}
\end{align}
which using \cite[2.321]{table_integrals},
simplifies to
\begin{align}
&=-\frac{1}{\Gamma(M_i)}\sum_{k=0}^{M_i-1}\frac{(-1)^{M_i-1-k} (M_i-1)!}{k!(-1)^{M_i-1-k+1}} \nonumber
\left(\frac{\beta_i(I+\sigma^2)}{P_iL(x_i)}\right)^ke^{-\left(\frac{\beta_i(I+\sigma^2)}{P_iL(x_i)}\right)} \nonumber\\
&\eqtext{(a)}e^{-\frac{\beta_i\sigma^2}{P_iL(x_i)}}\sum_{k=0}^{M_i-1}\left(\frac{\beta_i}{P_iL(x_i) }\right)^k\frac{1}{k!} \sum_{l=0}^k\binom{k}{l}(\sigma^2)^{k-l}I^le^{-\frac{\beta_iI}{P_iL(x_i)}} ,
\label{eq16}
\end{align}
where, (a) is obtained through
the binomial expansion.
Averaging \eqref{eq16} over interference results in
\begin{multline}
\label{cond_prob_averaged}
\mathbb{P}\left(\frac{P_ih_{x_i}L(x_i)}{I_{x_i}+\sigma^2} > \beta_i\right)
=e^{-\frac{\beta_i\sigma^2}{P_iL(x_i)}}\sum_{k=0}^{M_i-1}\frac{(\frac{\beta_i}{P_iL(x_i)})^k}{k!} \\
\times \sum_{l=0}^k\binom{k}{l}(\sigma^2)^{k-l}\mathbb{E}_I\left[I^le^{-\frac{\beta_iI}{P_iL(x_i)}}\right]  
\, .
\end{multline}
We simplify $\mathbb{E}_I\left[I^le^{-\frac{\beta_iI}{P_iL(x_i)}}\right]$ as follows
\begin{align}
\mathbb{E}_I\left[I^le^{-sI}\right]&=\int_{-\infty}^{\infty}y^le^{-sy}f_{I}(y)dy
=\mathcal{L}\left\{ y^lf_{I}(y) \right\}(s) \nonumber
\\
&=(-1)^l\frac{d^l\, \mathcal{L}\left\{ f_{I}(y) \right\}(s)}{ds^l} 
=(-1)^l\frac{d^l}{ds^l}\frac{\left\{\mathbb{E}_I\left[e^{-sI}\right]\right\}}{ } 
\, ,
\label{231}
\end{align}
where $\mathcal{L}\{.\}$ is the Laplace transform.
Substituting \eqref{231} in \eqref{cond_prob_averaged} 
and $s={\beta_i}/{(P_iL(x_i))}$
results in
\begin{align}
e^{-\frac{\beta_i\sigma^2}{P_iL(x_i)}}\sum_{k=0}^{M_i-1}\left(\frac{\beta_i}{P_iL(x_i) }\right)^k\frac{1}{k!} \sum_{l=0}^k\binom{k}{l}(\sigma^2)^{k-l}(-1)^l 
\frac{d^l}{ds^l}\frac{\left\{\mathbb{E}_I\left[e^{-sI}\right]\right\}}{ } \, .
\label{241}
\end{align}
An expression of
$\mathbb{E}_I\left[e^{-sI}\right]$ has been
expressed in \cite{downlink_mimo}
as follows
\begin{equation}
 \exp\lbrak{ -(s)^{2/\alpha} \!\! \sum_{m=1}^K\lambda_m(P_m)^{2/\alpha}}\!\! 
 \rbrak{\!\!  \sum_{p=1}^{M_m}\binom{M_m}{p} \frac{2\pi}{\alpha} \mathcal{B}(M_m-p+2/\alpha,p-2/\alpha)}
\end{equation}
where, $\mathcal{B}\brak{\cdot,\cdot}$ is the Beta function as given in \eqref{eqn_Ii} \cite[8.380]{table_integrals}.

The Faa Di Bruno formula \cite{encyclopedia_math}
can be used to obtain,
\begin{align}
\label{faa_bruno}
\frac{d^l}{ds^l}\frac{\left\{\mathbb{E}_I\left[e^{-sI}\right]\right\}}{ }=\sum_{r=0}^l f^{(r)}(g)B_{l,r}(g',g'',\dots,g^{l-r+1}) \, 
\end{align}
where,
\begin{equation}
f=e^x,\, g=-As^{2/\alpha},
\end{equation} 
such that their higher order derivatives are
\begin{equation}
\label{derv_fg}
f^{(r)}(g) = e^{-As^{2/\alpha}} , \,
g^{(t)} =-As^{\frac{2}{\alpha} -t}D_t
\end{equation}
and $A$, $D_t$, $
B_{l,r}\brak{g',\dots,g^{l-r+1}} $
are as expressed in \eqref{eqn_Ii}.
%
%
Bell polynomial in \eqref{faa_bruno} can be further simplified as,
\begin{align}
\label{bell_simplified}
B_{l,r}\brak{g',\dots,g^{l-r+1}} 
&=\sum\frac{l!}{j_1!j_2!\dots j_{l-r+1}!}\prod_{t=1}^{l-r+1}\left(\frac{-AD_ts^{\frac{2}{\alpha} -t}}{t!}\right)^{j_t} 
\nonumber
\\
&=\sum s^{\frac{2}{\alpha}(j_1+j_2+\dots+j_{l-r+1})}s^{-(j_1+2j_2+\dots+(l-r+1)j_{l-r+1})}
 \nonumber
 \\
&\qquad \times \frac{(-A)^{(j_1+j_2+\dots+j_{l-r+1})}l!}{j_1!j_2!\dots j_{l-r+1}!}\prod_{t=1}^{l-r+1}\left(\frac{D_t}{t!}\right)^{j_t} 
\nonumber
\\
&=(-A)^rs^{\frac{2r}{\alpha}-l}B_{l,r}(D_1,D_2,\dots,D_{l-r+1}).
\end{align}
Substituting \eqref{faa_bruno}, \eqref{derv_fg}, and \eqref{bell_simplified} in \eqref{241} results in
\begin{align}
\label{cond_prob}
\mathbb{P}&\left(\frac{P_ih_{x_i}L(x_i)}{I_{x_i}+\sigma^2} > \beta_i\right)
= \sum_{k=0}^{M_i-1}\frac{(\frac{\beta_i}{P_i})^k}{k!} \sum_{l=0}^k\binom{k}{l}(\sigma^2)^{k-l}(-1)^l
\nonumber
\\
&\quad \times \sum_{r=0}^l \left(\frac{\beta_i}{P_i}\right)^{\frac{2r}{\alpha}-l}(-A)^r B_{l,r}(D_1,D_2,\dots,D_{l-r+1})\nonumber
\\
& \quad \times  e^{-\frac{\beta_i\sigma^2||x_i||^{\alpha}}{P_i}-A(\frac{\beta_i}{P_i})^{2/\alpha}||x_i||^2} ||x_i||^{\alpha k+ 2r-\alpha l}
\, .
\end{align}
Using \eqref{cond_prob}, the probability of coverage in \eqref{general_cov} is expressed as
 \begin{align}
 P_c
&=\sum_{i=1}^K \lambda_i \int_{R^2}\sum_{k=0}^{M_i-1}\frac{(\frac{\beta_i}{P_i})^k}{k!} \sum_{l=0}^k\binom{k}{l}(\sigma^2)^{k-l}(-1)^l\sum_{r=0}^l \left(\frac{\beta_i}{P_i}\right)^{\frac{2r}{\alpha}-l}(-A)^r\nonumber
\\
& \quad \times  B_{l,r}(D_1,D_2,\dots,D_{l-r+1})e^{-\frac{\beta_i\sigma^2||x_i||^{\alpha}}{P_i}-A(\frac{\beta_i}{P_i})^{2/\alpha}||x_i||^2} ||x_i||^{\alpha k+ 2r-\alpha l}dx_i 
\label{311}
\end{align}
Converting \eqref{311} to polar form
along with transformation of variable 
results in
\begin{align}
P_c&=\sum_{i=1}^K \pi\lambda_iP_i^{2/\alpha}\beta_i^{-2/\alpha} \sum_{k=0}^{M_i-1}\frac{1}{k!} \sum_{l=0}^k\binom{k}{l}(\sigma^2)^{k-l}(-1)^l
\sum_{r=0}^l (-A)^r\nonumber\\
& \quad \times B_{l,r}(D_1,D_2,\dots,D_{l-r+1})\int_{0}^{\infty} e^{-\sigma^2t^{\alpha/2}-At} t^{r+\frac{\alpha}{2}(k- l)}dt \, .
\label{321}
\end{align}
Substituting the result obtained in \eqref{pl_integral_eqn} in \eqref{321}
results in \eqref{nak_cov_eqn}.
This completes the proof of Theorem~\ref{nak_cov_thm}.
\hfill \rule{1.5ex}{1.5ex}

\section{Proof of Theorem IV.1}
\label{rate_app}
Substituting $P_c$ from \eqref{nak_cov_eqn} in \eqref{ratio_Pc} we have,
\begin{equation}
\label{ratio_Pc_sub}
\mathbb{P}(X>y|\mathbf{C}(\{\beta_i\})) 
= \frac{\sum_{i=1}^K \pi\lambda_iP_i^{2/\alpha}max(y,\beta_i)^{-2/\alpha} \mathcal{I}_i }{\sum_{i=1}^K \pi\lambda_iP_i^{2/\alpha}\beta_i^{-2/\alpha} \mathcal{I}_i } \, .
\end{equation}
Using  \eqref{ratio_Pc_sub} in \eqref{rate_integral} gives,
\begin{align}
R&=\int_{0}^{\infty}\frac{1}{(1+y)}\frac{\sum_{i=1}^K \pi\lambda_iP_i^{2/\alpha}\beta_i^{-2/\alpha}\mathcal{I}_i \left(\beta_i^{2/\alpha} max(y,\beta_i)^{-2/\alpha}\right)}{\sum_{i=1}^K \pi\lambda_iP_i^{2/\alpha}\beta_i^{-2/\alpha} \mathcal{I}_i }dy
\\
&=\frac{\sum_{i=1}^K \pi\lambda_iP_i^{2/\alpha} \beta_i^{-2/\alpha} \mathcal{I}_i\mathcal{A}_i}{\sum_{i=1}^K \pi\lambda_iP_i^{2/\alpha}\beta_i^{-2/\alpha} \mathcal{I}_i } \nonumber \, ,
\end{align}
where,
\begin{align}
\mathcal{A}_i&=\int_{0}^{\infty}\frac{max(\beta_i,y)^{-2/\alpha}}{\beta_i^{-2/\alpha}(1+y)}dy \nonumber
\\
&=\int_{0}^{\beta_i}\frac{1}{1+y}dy+\frac{1}{\beta_i^{-2/\alpha}}\int_{\beta_i}^{\infty}\frac{y^{-2/\alpha}}{1+y}dy \nonumber
\\
&=\ln\left(1+\beta_i\right) +\frac{\alpha}{2} \ _2F_1\left(1,\frac{2}{\alpha},1+\frac{2}{\alpha};-\frac{1}{\beta_i}\right) \, ,
\end{align}
using the following in the second integral \cite[3.194]{table_integrals}
\begin{multline}
\int_u^{\infty}\frac{y^{\mu - 1}}{(1+\beta y)^v} dy=\frac{u^{\mu-v}}{\beta^{v}(v-\mu)}\  _2F_1\left(v,v-\mu,v-\mu +1;-\frac{1}{\beta u}\right)\nonumber
\\ \qquad Re\{v\}>Re\{u\} \, .
\end{multline}
This completes the proof of Theorem~\ref{nak_rate_thm}.
\hfill \rule{1.5ex}{1.5ex}

%

\end{document}